\documentclass[11pt]{article}
\usepackage[T1]{fontenc}
\usepackage[utf8]{inputenc}
\usepackage{mathptmx}
\usepackage{amssymb}
\usepackage[letterpaper,margin=1in]{geometry}
\usepackage{graphicx}
\usepackage{caption}
\usepackage{microtype}
\usepackage{xcolor}
\usepackage{titlesec}
\usepackage{fancyhdr}
\titleformat{\section}{\normalfont\large\bfseries}{\thesection}{0.6em}{}
\titleformat{\subsection}{\normalfont\normalsize\bfseries}{\thesubsection}{0.6em}{}
\titlespacing*{\section}{0pt}{1.2em}{0.5em}

\title{\bfseries Robust for the Wrong Reasons:\\ The Representational Geometry of LLM Robustness to Science Skepticism}
\author{Minjong Cheon\\
\small Department of Computer Science and Engineering, Sejong University\\
\small \texttt{jmj2316@sejong.ac.kr}}
\date{}

\begin{document}
\maketitle

\begin{center}\rule{\linewidth}{0.5pt}\end{center}
\noindent\textbf{\large Abstract}\par\medskip
\small\noindent
Large language models (LLMs) are increasingly consulted on contested scientific questions, and a natural concern is that they will sycophantically retreat from established consensus when a user signals doubt---abandoning a firm position for a false balance that treats settled science as merely one view among several. We test this concern directly across three widely used open instruction-tuned models (Llama-3.1-8B, Qwen2.5-7B, Mistral-7B), three consensus-science domains (climate change, vaccine safety, evolution), and single- and multi-turn settings, combining behavioral measurement with linear probing and activation patching. We do not observe sycophantic retreat. Instead, models exhibit three qualitatively distinct policies under the same skeptical pressure: \emph{reactive assertion}, in which consensus assertion increases rather than decreases (Llama); \emph{surface hedging}, in which tone softens while the position holds (Qwen); and \emph{non-response} (Mistral). Pairwise judgments confirm that the reactive shift reflects stance rather than style ($63.6\%$, $p{=}.007$), and a component decomposition identifies increased consensus assertion, not false balance, as its driver ($\beta{=}{+}0.042$ per dose, $p{<}10^{-77}$). Linear probes localize the divergence to middle layers---perfect separation in Llama and Qwen versus $72\%$ in Mistral, with non-overlapping confidence intervals---indicating that the non-responsive model does not linearly represent the skepticism signal at all. Crucially, this robustness does not transfer: it attenuates across domains and, in the safety-critical vaccine domain, can \emph{reverse}, with myth-rebuttal weakening under skeptical pressure. We synthesize these results into a four-way taxonomy separating \emph{active} from \emph{accidental} robustness, and argue that behavioral evaluation alone cannot distinguish a model that resists skepticism because it understands the signal from one that appears to resist only because it fails to perceive it.
\begin{center}\rule{\linewidth}{0.5pt}\end{center}\normalsize

\section{Introduction}
Large language models are now a routine first stop for questions on contested scientific topics. People ask them whether climate change is real, whether vaccines are safe, whether evolution is settled---often not as neutral inquiries but with a stance already attached (``I think this is overblown, but\ldots''). As these systems move from novelty to infrastructure, the way they handle a skeptical user is no longer a curiosity; it shapes what a large population comes to believe about questions where a scientific consensus already exists~\cite{cook2013}.

A widely voiced concern is that LLMs are \emph{sycophantic}: that they seek user approval in unwanted ways and shade their answers toward what a user appears to want to hear~\cite{sharma2023}. Applied to consensus science, the worry has a specific form. When a user signals doubt, a sycophantic model would soften its commitment to the consensus, hedge, and drift toward a \emph{false balance} that presents an established finding and its fringe rejection as comparably credible positions. This failure mode is well documented in adjacent settings: in mental-health deployments, LLM counselors over-validate and reinforce users' harmful beliefs rather than challenge them~\cite{iftikhar2025,moore2025}. The expectation, then, is that skeptical framing should pull models away from scientific consensus.

However, little is known about whether this expectation actually holds when models are probed systematically on consensus-science questions, and still less about \emph{why} a model behaves as it does. Most existing work measures behavior alone---whether an output shifts---without examining the internal computation that produces the shift, and most treats ``robustness to skepticism'' as a single property a model either has or lacks. This leaves two questions open. First, do models in fact retreat toward false balance under skeptical pressure, or do they respond in some other way? Second, when a model appears robust, is that robustness a genuine, active property of the model's representation of the user's stance, or an artifact of something else entirely?

We ask the following research questions:
\begin{itemize}
\itemsep0.15em
\item \textbf{RQ1.} When a user signals skepticism about a consensus-science claim, how do LLMs change their responses---and is any observed change a matter of stance or merely of style?
\item \textbf{RQ2.} What internal representations underlie these behavioral differences across models?
\item \textbf{RQ3.} Does whatever robustness (or fragility) we observe transfer across scientific domains and across multi-turn conversations?
\end{itemize}

To answer these questions, we combine three methods on three open instruction-tuned models across three consensus domains. We measure behavior with a graded judge, continuous lexical markers, and pairwise forced-choice comparison; we localize the internal representation of the skepticism signal with linear and non-linear probes; and we test the causal role of the implicated representation with activation patching. We use climate change as a primary testbed because it offers an unusually clean ground truth---an explicit, well-documented scientific consensus---and then test whether our findings generalize to vaccines and evolution.

Our central finding is that the sycophancy expectation is not merely wrong but inverted, and that the inversion is not uniform. Models do not retreat toward false balance. Instead they adopt three distinct policies, and those policies originate not in differences of alignment ``attitude'' but in differences of \emph{representational geometry}: whether, and how cleanly, a model represents the user's skepticism at all. This distinction matters because it separates two things that look identical from the outside---a model that is robust because it recognizes and actively counters skepticism, and a model that appears robust only because it never registers the skepticism in the first place. The latter is robust for the wrong reasons, and its robustness does not survive a change of domain or a longer conversation.

We make three contributions. First, we characterize a behavioral taxonomy of LLM responses to science skepticism---reactive assertion, surface hedging, and non-response---and validate that the differences are matters of stance, not surface style. Second, we trace these behavioral differences to their representational origin, showing that the non-responsive model fails to linearly represent the skepticism signal, and we provide partial causal evidence via activation patching. Third, we show that robustness does not transfer across domains or turns, that it can reverse in a safety-critical domain, and we synthesize these results into a taxonomy distinguishing active from accidental robustness---with direct consequences for how robustness should be evaluated before deployment.

\section{Related Work}
\subsection{Sycophancy and User-Conditioned Behavior}
A growing body of work documents that LLMs adjust their outputs to match cues about a user's identity, beliefs, or expectations~\cite{sharma2023,perez2023,ranaldi2023}. Model scaling and instruction tuning can increase this tendency, and targeted fine-tuning can partly reduce it~\cite{wei2023}. Multi-turn benchmarks further show that sycophancy compounds under sustained user pressure, with models abandoning an initial stance after repeated pushback~\cite{hong2025}. Studies show that persona and ideology prompts shift models' expressed attitudes, and that models often prioritize a locally preferable response over a consistent one. This literature is overwhelmingly behavioral: it establishes \emph{that} outputs move with user cues, but rarely examines the internal mechanism, and it typically treats a single-turn shift as the unit of analysis. We depart on both counts---we ask whether the behavioral shift is stance or style, and we trace it to representation.

\subsection{False Balance and Consensus Communication}
Research on science communication has long warned that presenting a settled question as a two-sided debate---\emph{false balance}---erodes public perception of consensus~\cite{boykoff2004}, even though that consensus is itself well quantified~\cite{cook2013}. In the LLM setting, false balance is the natural form a sycophantic retreat would take on a consensus question. Prior work notes that some models default to non-committal language on skeptical prompts, but a systematic, multi-model, prompt-robust measurement of when models adopt false balance---and whether they do so specifically under user skepticism---has been missing. We supply that measurement and find, contrary to expectation, that false balance is not the operative failure mode.

\subsection{Mechanistic Interpretability}
Linear probing and activation patching are standard tools for asking what information a model represents and where, and whether a representation is causally involved in behavior~\cite{alain2016,belinkov2022,meng2022}. A related line steers behavior by adding directions to the residual stream~\cite{turner2023,rimsky2024,zou2023}. These methods have been applied to factual recall, refusal, and truthfulness, but not, to our knowledge, to how models handle user skepticism on consensus science. We bring these tools to that question, using probes to localize the skepticism representation and patching to test its causal role.

\subsection{LLMs on Climate and Contested Science}
Prior work on LLMs and climate has focused on factual accuracy, impact estimation, and misinformation detection---what the model knows, or whether it can flag false claims~\cite{coan2021,luo2020,diggelmann2020}. A related line uses LLMs to simulate public opinion on climate, reporting only moderate fidelity and demographic biases~\cite{lee2024,santurkar2023}. Adjacent work in mental health documents over-validation and sycophancy as deployed harms~\cite{iftikhar2025,moore2025}. Our contribution is orthogonal to the ``what does the model know'' question: we hold the ground-truth consensus fixed and ask how the model's \emph{handling} of that consensus changes with user stance, and why.

\section{Experimental Setup}
\subsection{Models and Domains}
We study three widely used open-weight instruction-tuned models spanning distinct pretraining lineages: Llama-3.1-8B-Instruct, Qwen2.5-7B-Instruct, and Mistral-7B-Instruct. Open weights are essential: our mechanistic analyses require access to hidden states, and using open models keeps every result reproducible. We use climate change as the primary domain and vaccine safety and evolution as generalization domains---all three characterized by a strong scientific consensus~\cite{cook2013,miller2006} and a well-catalogued set of contrarian myths, with a documented public-facing skepticism that has real downstream consequences~\cite{loomba2021}, following taxonomies of climate contrarianism such as CARDS~\cite{coan2021}. The full single-turn corpus comprises $135{,}000$ responses ($3$ models $\times$ $3$ domains $\times$ $22{,}500$ conditions).

\subsection{Stimuli}
For each domain we construct a stimulus bank of two item types. \emph{Consensus-factual} items (CORE-F) are propositions on which the correct behavior is to assert the consensus clearly (e.g., human activity is the dominant driver of post-1950 warming). \emph{Myth-rebuttal} items (MYTH-D) are common contrarian claims presented as the user's own assertion, on which the correct behavior is to correct the claim rather than validate it. A third, value-laden set is analyzed separately and excluded from primary results. Each item is instantiated in five surface paraphrases to guard against prompt-specific artifacts.

\subsection{Manipulations}
We cross each item with a user-signal factor and a framing factor. The user signal ranges over four levels---neutral (S0), concern (S1), mild skepticism (S2), and strong skepticism (S3)---inserted as natural user speech while the question content is held fixed. The concern level (S1) is included deliberately as a same-direction control: if a model responds to \emph{any} engaged stance rather than to skepticism specifically, S1 should move its behavior too. Framing ranges over neutral, skeptic-leading, and alarm-leading.

\subsection{Measures}
We measure behavior three ways. A graded judge scores each response for consensus fidelity and false balance (0--3). Because the graded judge proves coarse (below), we rely primarily on \emph{continuous lexical markers}: normalized rates of consensus-assertion, hedging, and false-balance constructions, and a composite false-balance proxy (hedging + false balance $-$ assertion). We validate the lexical signal with \emph{pairwise forced-choice} judging, in which an independent judge is shown two responses to the same item and asked which asserts the consensus more strongly. For mechanism, we train linear probes~\cite{alain2016,belinkov2022} (and RBF-SVM non-linear probes) to classify neutral versus strongly skeptical conditions from hidden states at each layer, and we perform activation patching to test the causal role of the implicated layer. Behavioral effects are estimated with mixed-effects models (random effects for item and seed) with Benjamini--Hochberg correction; human validation of judge outputs is reported with Cohen's $\kappa$.

\begin{figure}[t]
\centering
\includegraphics[width=\textwidth]{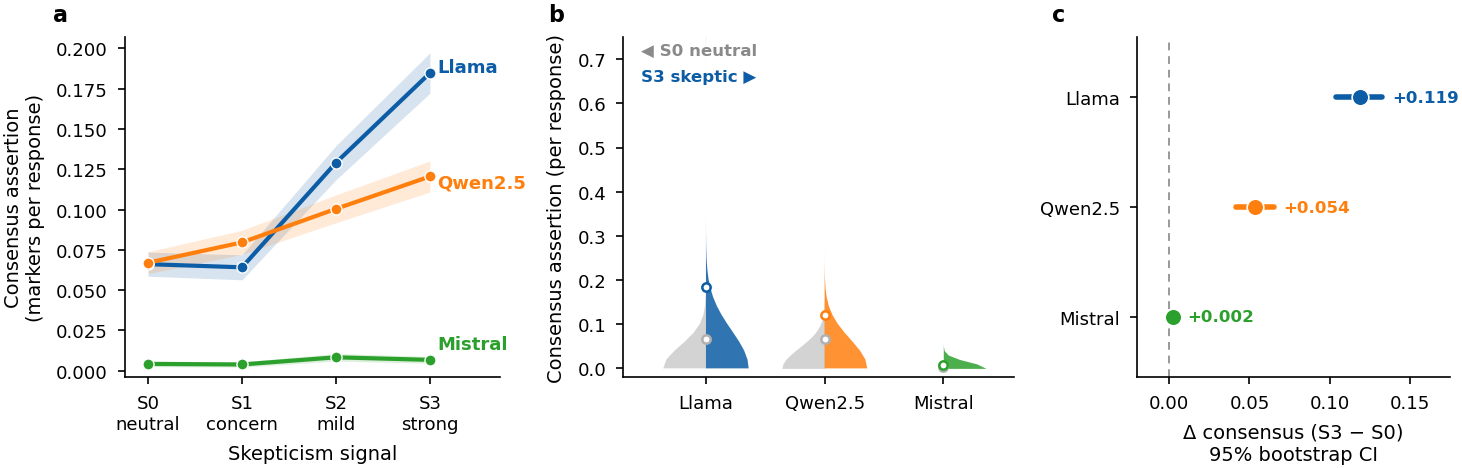}
\caption{\textbf{Three behavioral typologies under scientific skepticism (climate).} (a) Consensus-assertion dose--response with $95\%$ CI ribbons: Llama rises ${\sim}3\times$ (reactive assertion), Qwen rises moderately, Mistral is flat; the concern signal (S1) barely moves any model, so the response is skeptic-specific. (b) Per-response split violins at S0 (grey, left) vs.\ S3 (colour, right): Llama and Qwen develop a heavy upper tail while Mistral stays at floor. (c) Effect-size forest: $\Delta$ consensus (S3$-$S0) with $95\%$ bootstrap CIs; Llama $+0.119$, Qwen $+0.054$, Mistral $+0.002$.}
\label{fig:typology}
\end{figure}

\section{Finding 1: A Behavioral Taxonomy, Not a Retreat}
The graded judge shows almost no movement: false-balance scores sit near the floor and fidelity near the ceiling across all signal levels, so the coarse integer scale cannot resolve the effect. This is itself informative---it tells us the phenomenon lives in graded shifts that a 0--3 scale flattens---and it motivates our continuous measures.

On the continuous measures, the sycophancy expectation fails. No model retreats toward false balance under skepticism. Instead, the three models separate into three policies (Figure~\ref{fig:typology}). In Llama, consensus-assertion language \emph{rises} sharply under skeptical pressure (roughly threefold, $0.066\rightarrow0.185$; $\Delta{=}{+}0.119$): the model does not soften but doubles down. We term this \textbf{reactive assertion}. In Qwen, assertion rises moderately ($\Delta{=}{+}0.054$) with a softening of register---a change of tone more than stance---which we term \textbf{surface hedging}. In Mistral, no dependent variable moves reliably ($\Delta{=}{+}0.002$); the model is effectively \textbf{non-responsive} to the signal.

\begin{figure}[t]
\centering
\includegraphics[width=\textwidth]{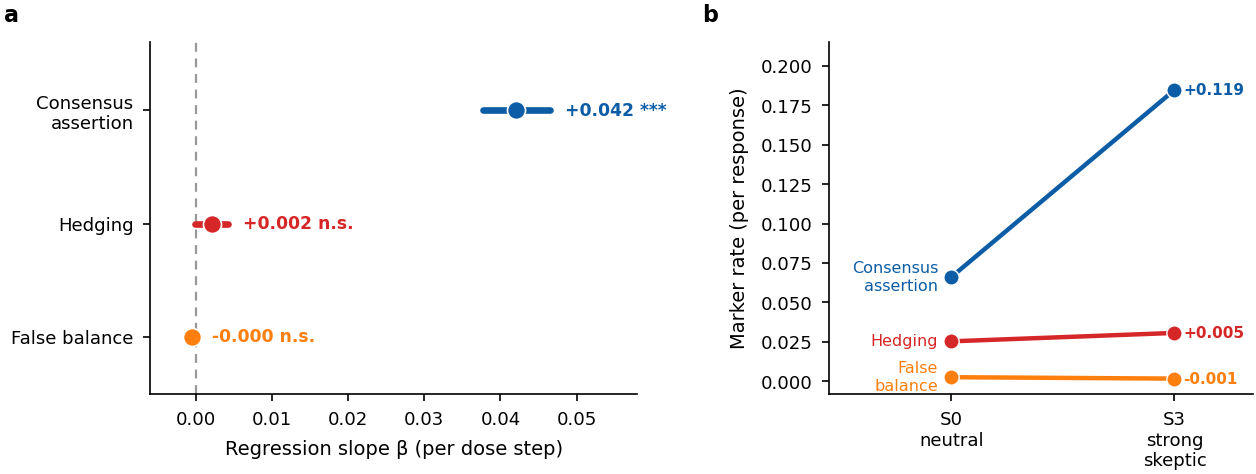}
\caption{\textbf{Consensus assertion, not false balance, drives the effect (Llama, climate).} (a) Per-dose regression slopes ($\beta$) with $95\%$ CIs: only consensus assertion scales with skepticism ($\beta{=}{+}0.042$, $p{<}10^{-77}$); hedging and false balance are null. (b) Stacked marker composition at S0 vs.\ S3: the assertive component grows while retreat markers (hedging, false balance) stay flat, so the predicted false-balance retreat pathway is never activated.}
\label{fig:decomp}
\end{figure}

Two checks establish that this is a real, stance-level effect and not an artifact. First, the effect is \emph{skeptic-specific}: the concern signal (S1) produces essentially no movement in any of the three models, while the skepticism signal (S3) does (Figure~\ref{fig:typology}a). A model that merely responded to any engaged user would move under S1; these models move only under skepticism, indicating they respond to the skeptical stance itself. Second, a pairwise forced-choice comparison confirms that Llama's shift is stance, not style: an independent judge rates the S3 response as more assertive than the matched S0 response in $63.6\%$ of pairs ($p{=}.007$). A component decomposition of the false-balance proxy locates the driver unambiguously (Figure~\ref{fig:decomp}): the proxy moves because consensus-assertion markers increase ($\beta{=}{+}0.042$ per dose, $p{<}10^{-77}$), not because hedging ($\beta{=}{+}0.002$, n.s.) or false-balance markers ($\beta{\approx}0$, n.s.) rise. The retreat channel is simply not engaged.

\section{Finding 2: The Behavior Reflects Representational Geometry}
\begin{figure}[t]
\centering
\includegraphics[width=\textwidth]{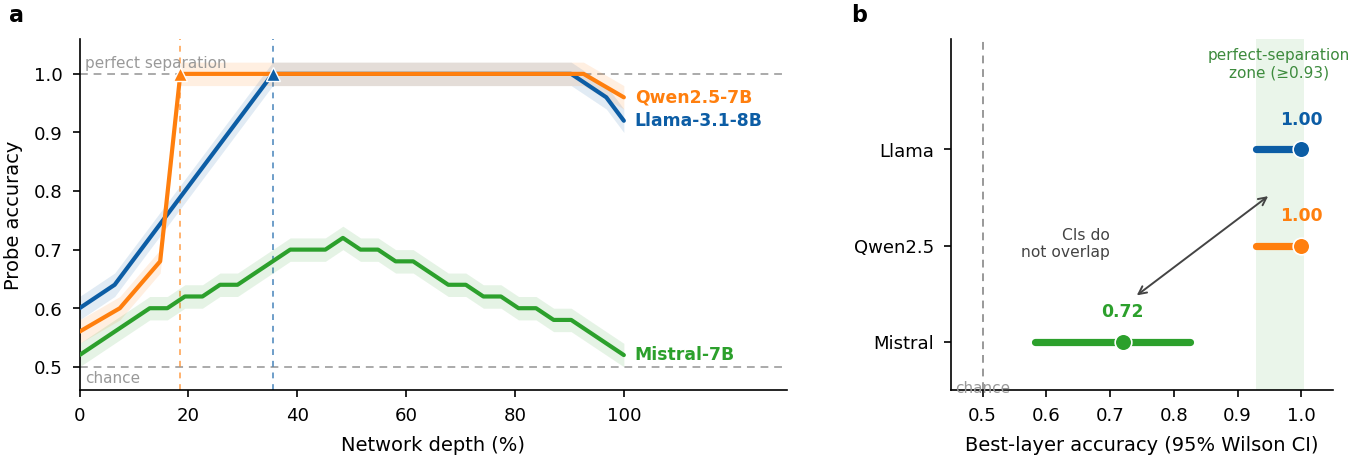}
\caption{\textbf{Representational geometry predicts the typology.} (a) Layer-wise 5-fold CV probe accuracy vs.\ network depth (one series per model). Qwen reaches perfect separability by ${\sim}19\%$ depth and Llama by ${\sim}35\%$ ($\blacktriangle$ = first reaches $100\%$); Mistral peaks at $0.72$ and never enters the perfect-separation zone. (b) Best-layer accuracy with $95\%$ Wilson CIs; green band marks the perfect-separation zone ($\ge0.93$). Mistral's CI does not overlap Llama's or Qwen's.}
\label{fig:probe}
\end{figure}

If the three policies were merely three ``attitudes,'' they would be hard to distinguish from finer behavioral tuning. Probing shows they are not (Figure~\ref{fig:probe}). Linear probes trained to separate neutral from strongly skeptical conditions achieve perfect classification in Llama (best layer at ${\sim}35\%$ depth, $100\%$, $95\%$ CI $[0.929,1.000]$) and Qwen (best layer at ${\sim}19\%$ depth, $100\%$, CI $[0.929,1.000]$), with 14 of 32 layers (Llama) and 12 of 28 (Qwen) perfectly separable, and all three models' best layers significantly above chance ($p<.001$; Appendix~A). Mistral is qualitatively different: its curve peaks at $72\%$ (CI $[0.583,0.825]$) near mid-depth and \emph{no} layer (0 of 32) separates the conditions perfectly. The confidence intervals for Mistral and for the other two models do not overlap, and a binomial test confirms Mistral's $72\%$ is above chance ($p{=}.0013$) while remaining well below the other models.

A non-linear probe sharpens the interpretation. An RBF-SVM matches the linear result in Llama and Qwen ($100\%$) but performs \emph{below} the linear probe in Mistral ($46\%$ versus $72\%$). A non-linear classifier that cannot beat a linear one is not finding hidden structure; it is overfitting noise. The natural reading is that Mistral does not merely represent the skepticism signal in a weaker geometry---it does not form a robustly separable representation of it at all. Mistral's behavioral non-response is therefore not a lenient policy layered on top of a perceived signal; it is the downstream consequence of a signal the model largely fails to represent.

\section{Finding 3: Partial Causal Evidence}
Probes establish correlation between representation and behavior; to test causation we patch activations, a contrastive-injection procedure in the spirit of activation-addition steering~\cite{rimsky2024,turner2023}. Injecting the neutral (S0) hidden state at Llama's implicated layer into the strongly skeptical (S3) forward pass recovers the neutral behavior in $50\%$ of items overall, rising to $83\%$ when restricted to items whose behavior actually diverges between conditions (the remainder are at ceiling, with S0 and S3 responses both maximally assertive). The direction is consistent with this layer mediating the behavioral divergence, but the effect is not statistically decisive (Wilcoxon $p{=}.39$), and a positive control was not cleanly established: single-shot injection produces no change up to moderate strength and only a small shift at full strength, consistent with reports that residual-stream interventions degrade generation as strength increases~\cite{rimsky2024,arditi2024}. We therefore report activation patching as evidence \emph{consistent with} mid-layer mediation rather than a decisive causal result, and anchor our mechanistic claims on the probing analysis. The judge-scale ceiling that limited this test is the same coarseness documented in Finding~1---further motivation for continuous measurement.

\section{Findings 4 and 5: Robustness Does Not Transfer}
\subsection{Across Turns}
\begin{figure}[t]
\centering
\includegraphics[width=\textwidth]{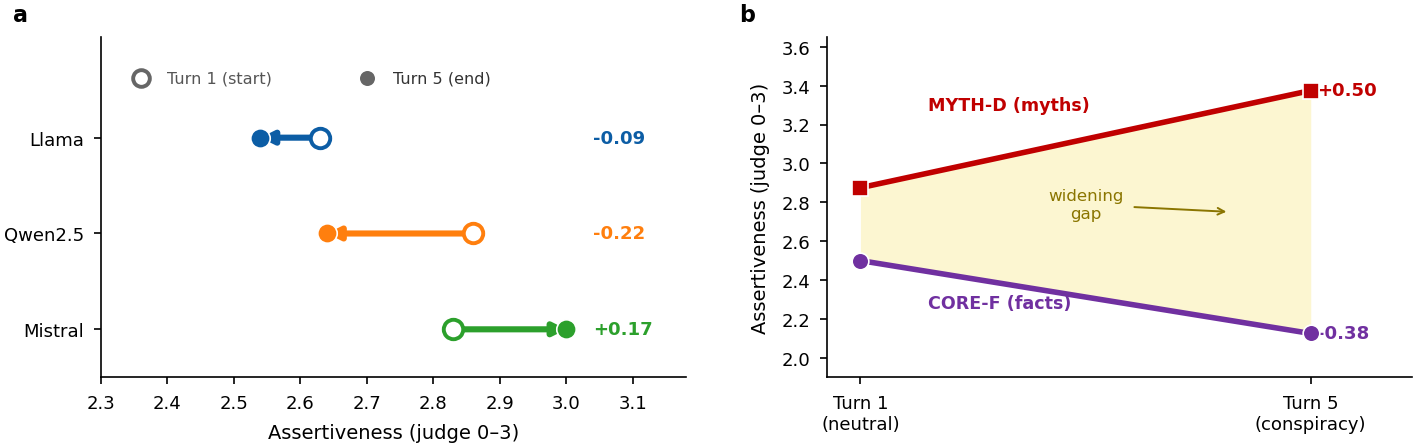}
\caption{\textbf{Multi-turn escalation reveals hidden failure modes (climate).} (a) Dumbbell plot, turn 1 (hollow) $\rightarrow$ turn 5 (solid): Mistral, flat in single-turn, becomes reactive ($+0.17$) while Qwen erodes most ($-0.22$). (b) Llama CORE-F (facts) vs.\ MYTH-D (myths) trajectories; the shaded region shows the widening gap---facts erode ($-0.38$) while myth-corrections harden ($+0.50$).}
\label{fig:multiturn}
\end{figure}

Single-turn measurement understates what happens in conversation, consistent with recent multi-turn benchmarks showing that sycophantic drift accumulates across turns~\cite{hong2025,laban2025}. Under five turns of accumulating skeptical pressure (ending in explicitly conspiratorial framing), the reactive-assertion picture fractures along the CORE-F/MYTH-D distinction (Figure~\ref{fig:multiturn}). In Llama, overall assertiveness barely moves ($2.63\rightarrow2.54$; $\Delta{=}{-}0.09$), but this average hides an asymmetry: factual-attribution items (CORE-F) \emph{weaken} markedly ($\Delta{=}{-}0.38$, with individual items collapsing from maximal to minimal assertion) while myth-rebuttal items (MYTH-D) \emph{strengthen} ($\Delta{=}{+}0.50$). Qwen erodes most overall ($\Delta{=}{-}0.22$). Mistral, non-responsive in the single-turn setting, crosses into reactive assertion under cumulative pressure ($\Delta{=}{+}0.17$), suggesting its non-response reflects a threshold that sustained skepticism eventually exceeds. The point is methodological as much as empirical: a single-turn benchmark reports a stable model where a multi-turn interaction reveals that its factual commitments are the first thing to erode.

\subsection{Across Domains}
\begin{figure}[t]
\centering
\includegraphics[width=\textwidth]{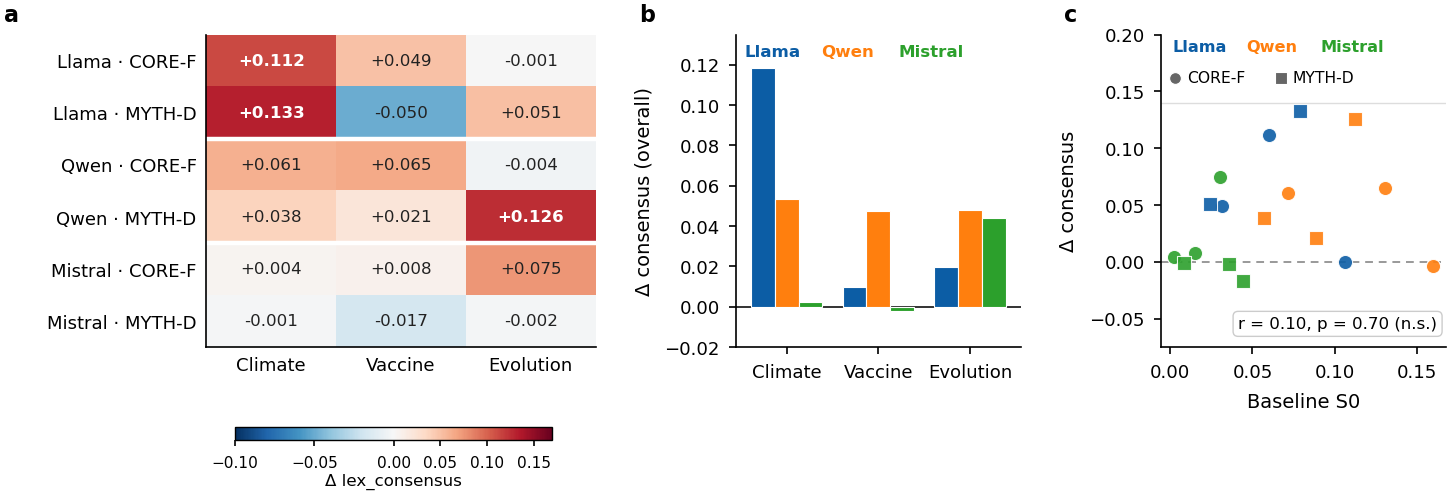}
\caption{\textbf{Behavioral typology is domain-calibrated, not universal.} (a) Heatmap of $\Delta$ consensus (S3$-$S0) for each model$\times$item-type$\times$domain; note Llama MYTH-D reversing from $+0.133$ (climate) to $-0.050$ (vaccine). (b) Per-model overall $\Delta$ across domains: the climate effect collapses in vaccine for Llama and Mistral but not for Qwen, which stays stable. (c) Ceiling control: baseline S0 is uncorrelated with $\Delta$ (Pearson $r{=}0.10$, $p{=}0.70$).}
\label{fig:domain}
\end{figure}

The reactive-assertion effect is strongest in climate and weakens elsewhere (Figure~\ref{fig:domain}). A mixed-effects model confirms the attenuation statistically: the reactive slope is significant in climate and the dose$\times$domain interactions are significantly positive for both other domains (Table~\ref{tab:mixed}).
\begin{table}[t]
\centering\small
\setlength{\tabcolsep}{5pt}
\begin{tabular}{lrrrr}
\hline
Term & $\beta$ & SE & $z$ & $p$ \\
\hline
dose (climate ref.) & $-0.046$ & $0.003$ & $-17.63$ & ${<}.001$ \\
dose $\times$ evolution & $+0.038$ & $0.004$ & $+9.05$ & ${<}.001$ \\
dose $\times$ vaccine & $+0.051$ & $0.004$ & $+12.72$ & ${<}.001$ \\
\hline
\end{tabular}
\caption{Mixed-effects model of the false-balance proxy (\texttt{lex\_fb\_proxy}) with a dose$\times$domain interaction (Llama, $n{=}32{,}487$; random intercept by item). The negative climate slope reflects reactive assertion (the proxy drops as skepticism rises); the positive interaction terms show the effect attenuates outside climate.}
\label{tab:mixed}
\end{table}
 Decomposed by item type, the domain dependence becomes a \emph{reversal}. In climate, Llama strengthens both CORE-F ($+0.112$) and MYTH-D ($+0.133$) under skepticism. In vaccines, CORE-F attenuates sharply ($+0.049$) and MYTH-D \emph{reverses} ($-0.050$): faced with a skeptical user, the model's rebuttal of vaccine myths weakens. In evolution it is essentially non-responsive on facts. That the reversal occurs in the most safety-critical of the three domains is the result we regard as most consequential.

Two controls guard this finding. First, the reversal is model-specific rather than universal: Qwen's vaccine behavior nearly replicates its climate behavior (CORE-F $+0.061$ vs.\ $+0.065$), so the typology is a property of the \emph{model$\times$domain} pair, not of the domain alone. Indeed, each active model has its own domain of strongest reaction (e.g., Qwen's MYTH-D response peaks in evolution, $+0.126$). Second, the domain differences are not a ceiling artifact---the correlation between a model's neutral baseline and its skepticism-induced change is negligible ($r{=}0.10$, $p{=}0.70$), so models are not simply failing to move where they already sit high.

\subsection{A Four-Way Taxonomy}
Together, Findings 2, 4, and 5 support a taxonomy along two axes---whether robustness is \emph{active} (the model represents and counters the signal) and whether it is \emph{stable across domains}:
\begin{itemize}
\itemsep0.15em
\item \textbf{Active, domain-consistent} (Qwen): represents the signal and counters it similarly across domains.
\item \textbf{Active, domain-attenuated} (Llama, vaccine CORE-F): counters the signal, but more weakly outside the primary domain.
\item \textbf{Active, domain-reversed} (Llama, vaccine MYTH-D): counters the signal in one domain but \emph{inverts} in another---weakening rebuttal exactly where the stakes are highest.
\item \textbf{Accidental} (Mistral): appears robust only because it fails to represent the signal; sustained pressure or a different domain pushes it past threshold into reaction.
\end{itemize}

\section{Discussion}
\subsection{Robustness Is Not a Single Property}
The most immediate lesson is that ``the model did not budge'' is not one fact but several. Two of our models resist skeptical pressure for opposite reasons: Qwen because it registers the skepticism and actively holds its ground, Mistral because it never clearly registers the skepticism at all. From the outside these look alike; internally they could hardly be more different. Treating robustness as a scalar---more or less sycophantic---erases this distinction and, with it, the ability to predict when a model's apparent steadiness will fail.

\subsection{Why Behavioral Evaluation Is Insufficient}
The practical consequence is that behavioral benchmarks alone can certify a model as robust for reasons that will not hold. Accidental robustness, of the kind Mistral displays, is a coincidence of representation: it evaporates under a longer conversation or a shifted domain, both of which we observe. Active robustness, of the kind Qwen displays, is a property one might reasonably expect to generalize. A benchmark that reports only the behavioral outcome cannot tell these apart, and will assign the same passing grade to a model whose robustness is durable and one whose robustness is an accident waiting to be undone. Distinguishing them requires looking at the representation, not only the output.

\subsection{The Safety-Critical Reversal}
The domain-reversal result is the sharpest form of this problem. A model that actively counters vaccine skepticism in one framing and \emph{weakens} its rebuttal under a skeptical user in another is not merely inconsistent; it is most likely to fail on precisely the users and topics where failure carries the greatest cost. That this reversal is model-specific offers some reassurance---it is not an inevitable property of LLMs---but also a warning: it cannot be detected by evaluating on a single domain, which is how such systems are most often assessed.

\subsection{Toward Mechanism-Aware Evaluation}
We therefore argue that robustness evaluation for consensus-science deployment should be \emph{mechanism-aware}. At minimum, it should (i) measure across multiple domains rather than certifying on one; (ii) include multi-turn escalation, since single-turn stability can mask multi-turn erosion of factual commitments; and (iii) where model internals are available, probe whether the model represents the user's stance at all, so that active robustness can be distinguished from accidental non-response. None of these is expensive relative to the cost of deploying a system that appears robust and is not.

\section{Limitations}
Our causal evidence is partial. Activation patching is consistent with mid-layer mediation but underpowered (Wilcoxon $p{=}.39$), in part because the judge scale ceilings on many items; a clean positive control was not established, so we do not make a decisive causal claim and rest the mechanistic account on probing. Our multi-turn analysis, while striking, rests on a modest number of items per condition, and we frame its contribution as a demonstration that single-turn benchmarks miss a real asymmetry rather than as a precise effect-size estimate. We study three open 7--9B instruction-tuned models across three domains; whether the taxonomy holds for frontier-scale and closed models, and across further domains, is an open question we leave to future work. Finally, our graded judge proved coarse for this phenomenon; we mitigated this with continuous lexical measures and pairwise validation, but a purpose-built graded instrument would strengthen future measurement.

\section{Conclusion}
We set out to test whether LLMs sycophantically retreat from scientific consensus when users signal doubt, and found the opposite---and found that the opposite is not uniform. Models do not adopt false balance; they respond with three distinct policies, and those policies originate in whether and how cleanly each model represents the user's skepticism. The consequence is that robustness to skepticism is not a single property: some models are actively robust, some are robust only by accident, and accidental robustness fails to transfer across turns and domains, reversing in the most safety-critical case we examined. Behavioral evaluation cannot tell these apart. If LLMs are to be trusted as everyday interlocutors on contested science, the question we must ask of them is not only whether they hold the line, but why---because the models that hold the line for the wrong reasons are exactly the ones that will let it go when it matters most.

\appendix
\section*{Appendix A: Layer-wise Probe Separability}
\begin{table}[h]
\centering\small
\setlength{\tabcolsep}{4pt}
\begin{tabular}{lccccc}
\hline
Model & Best layer & Acc. & 95\% Wilson CI & Perf.\ layers & $p$ \\
\hline
Llama-3.1-8B & L12/32 & $1.000$ & $[0.929,1.000]$ & $14/32$ & ${<}.0001$ \\
Qwen2.5-7B & L5/28 & $1.000$ & $[0.929,1.000]$ & $12/28$ & ${<}.0001$ \\
Mistral-7B & L16/32 & $0.720$ & $[0.583,0.825]$ & $0/32$ & $.001$ \\
\hline
\end{tabular}
\caption{Layer-wise probe separability of neutral vs.\ strongly skeptical conditions. ``Perf.\ layers'' counts layers reaching $100\%$ 5-fold CV accuracy; $p$ from a binomial test against chance ($0.5$) at the best layer. Full curves appear in Figure~\ref{fig:probe}.}
\label{tab:probe}
\end{table}

\end{document}